\documentclass[sigconf]{acmart}
\AtBeginDocument{%
  }

\setcopyright{acmlicensed}
\copyrightyear{2025}
\acmYear{2025}
\acmDOI{XXXXXXX.XXXXXXX}
\acmConference[EASE 2025]{The 29th International Conference on Evaluation and Assessment in Software Engineering}{17–20 June, 2025}{Istanbul, Turkey}

\usepackage{fancyvrb}
\usepackage{float}
\usepackage{graphicx}
\usepackage{enumitem}
\usepackage{tabularx}
\usepackage{booktabs}

\floatstyle{plain}
\newfloat{listing}{htbp}{lop}
\floatname{listing}{Listing}

\DefineVerbatimEnvironment{MyVerbatim}{Verbatim}{
  frame=single, 
  fontsize=\small, 
  numbers=left
}

\begin{document}

\title{Testing SSD Firmware with State Data-Aware Fuzzing: Accelerating Coverage in Nondeterministic I/O Environments}

\author{Gangho Yoon}
\orcid{0009-0009-6186-3065}
\affiliation{
  \institution{Sungkyunkwan University}
  \city{Suwon-si}
  \country{Republic of Korea}
}

\affiliation{
  \institution{Samsung Institute of Technology}
  \city{Yongin-si}
  \country{Republic of Korea}
}	
\email{yunkh21@gmail.com}

\author{Eunseok Lee}
\orcid{0000-0002-6557-8087}
\affiliation{%
  \institution{Sungkyunkwan University}
  \city{Suwon-si}
  \country{Republic of Korea}
}
\email{leees@skku.edu}

\renewcommand{\shortauthors}{Gangho Yoon and Eunseok Lee}

\begin{abstract}
Solid-State Drive (SSD) firmware manages complex internal states, including flash memory maintenance. Due to nondeterministic I/O operations, traditional testing methods struggle to rapidly achieve coverage of firmware code areas that require extensive I/O accumulation. To address this challenge, we propose a state data-aware fuzzing approach that leverages SSD firmware’s internal state to guide input generation under nondeterministic I/O conditions and accelerate coverage discovery. Our experiments with an open-source SSD firmware emulator show that the proposed method achieves the same firmware test coverage as a state-of-the-art coverage-based fuzzer (AFL++) while requiring approximately 67\% fewer commands, without reducing the number of crashes or hangs detected. Moreover, we extend our experiments by incorporating various I/O commands beyond basic write/read operations to reflect real user scenarios, and we confirm that our strategy remains effective even for multiple types of I/O tests. We further validate the effectiveness of state data-aware fuzzing for firmware testing under I/O environments and suggest that this approach can be extended to other storage firmware or threshold-based embedded systems in the future.
\end{abstract}

\ccsdesc[500]{Software and its engineering~Software testing and debugging}
\ccsdesc[300]{Software verification and validation}

\keywords{SSD Firmware, Fuzzing, Coverage-Based Testing, Nondeterministic I/O, Threshold}

\maketitle

\section{INTRODUCTION}
High-performance SSDs play a crucial role in modern computing systems and are widely used in environments ranging from data centers to high-end workstations. SSD firmware performs maintenance functions such as garbage collection (GC), wear-leveling, and error correction (ECC). Defects in this internal logic can lead to data corruption or performance degradation. Thus, verifying the reliability of SSD firmware is essential.

In practice, reliability is ensured during early firmware development by quickly identifying and resolving defects \cite{popereshnyak2019modeling}. However, algorithms that only activate after prolonged testing cannot be fully exercised within normal verification periods; thus, speeding up verification directly reduces quality-assurance costs \cite{li2015use}. Moreover, certain internal behaviors may remain undiscovered during unit tests but only emerge as defects when large numbers of commands are accumulated under real user scenarios \cite{strandberg2020intermittently}. Consequently, there is a need for a testing strategy that quickly achieves high code coverage under realistic user conditions. Designing test cases manually demands enormous time and effort, while randomly executing various SSD commands is also inefficient. Therefore, an automated test input generation method that considers the characteristics of SSD firmware is needed. Such an approach can accelerate the development process and yield more reliable results.

Coverage-based fuzzing has been shown to be extremely effective at testing software by injecting a large number of inputs into a program and leveraging execution-path feedback to guide test exploration intelligently \cite{serebryany2015libfuzzer, zalewski_afl}. Tools like AFL instrument the target program to trace which code path each input exercises, then save any input that reaches a new path as a seed \cite{zalewski_afl}. These techniques assume a deterministic relationship between inputs and coverage when testing a program \cite{meng2025aflnet}. However, testing systems that depend on hardware, such as firmware, pose additional challenges. Even the same input can yield different internal execution paths on each run, and such nondeterminism introduces noise into coverage feedback \cite{myung2022mundofuzz,vinesh2020confuzz}. This noise complicates the fuzzer’s judgment of valid inputs, ultimately hindering new path exploration \cite{meng2025aflnet}. Furthermore, hardware-related maintenance operations (e.g., GC, ECC) are generally not triggered by a single test input but only come into effect after hundreds of thousands of accumulated test inputs \cite{bux2014valid}. When each test input is processed in isolation, the relationship between commands and the firmware’s internal processes cannot be exposed, rendering it indistinguishable from random fuzzing. As a result, maintenance routines might never be executed before testing finishes, leaving a substantial portion of code untested. For example, some studies use reinforcement learning together with graph models to track SSD firmware’s long-term I/O state changes and boost coverage \cite{lee2023graft}. Such results highlight the importance of approaches that directly target maintenance routines and prolonged operations.

To solve these issues, this study proposes a state data-aware fuzzing approach specialized for SSD firmware. Our core idea is that the fuzzer recognizes changes in device state and reuses recently triggered inputs that caused those changes. The fuzzer stores sequences that induced such changes and replays them according to the current SSD memory usage state. This allows us to quickly achieve higher coverage in the firmware’s internal operations, a level of coverage that random searches alone would struggle to reach, and continue exploration.
 
\begin{figure*}[htbp]
  \centering
  \includegraphics[width=\textwidth]{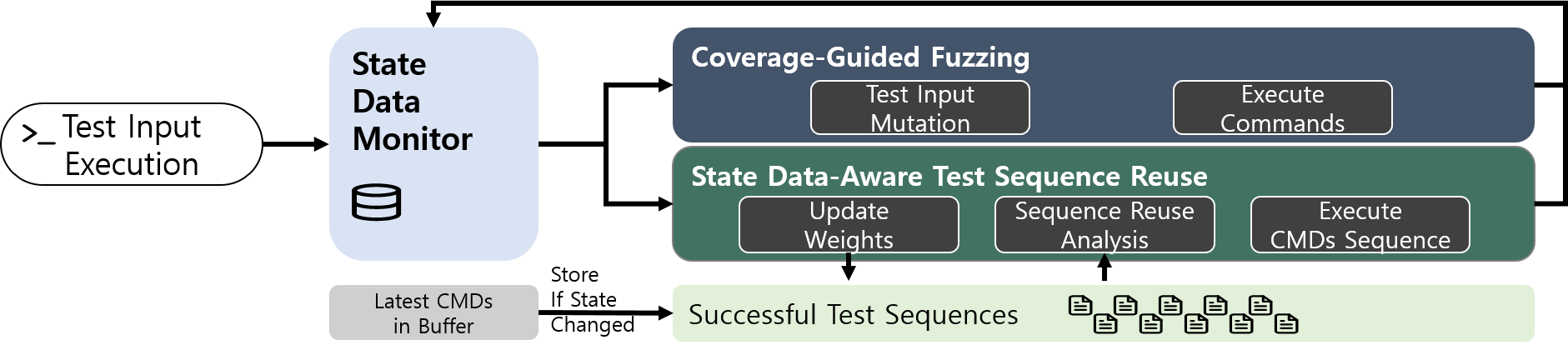}
  \caption{State Data-Aware Fuzzer Overview.}
  \Description{Overview of the proposed fuzzing framework}
\end{figure*}

\section{BACKGROUND AND RELATED WORK}
\paragraph{\textbf{SSD Firmware’s State-Based and Nondeterministic Behavior}}

Due to the nature of flash memory management, the Flash Translation Layer (FTL) maps logical addresses to physical pages, and as write/erase operations accumulate, metadata states continuously change. Such mechanisms \cite{nisbet2019trim,bux2014valid} only manifest after a sufficient internal state change has accumulated.

Moreover, the SSD controller handles parallel flash channels and asynchronous I/O requests simultaneously, so even the same input sequence can follow a different execution path depending on the device state or progress of background tasks. These frequent nondeterministic events mean that repeating the same input does not guarantee the same path every run.

\paragraph{\textbf{Threshold-Based Maintenance Logic Requiring I/O Accumulation}}
Key maintenance routines in SSD firmware (e.g., GC, wear-leveling, read disturb) are not triggered by a single command but require hundreds of thousands of I/O accumulations before manifesting. For example, even meticulously crafted test cases focusing on GC have been reported to require more than 258,000 I/Os to reach close to 100\% coverage \cite{lee2023graft}. It is also well known that sufficient valid pages must be consumed before the algorithm operates \cite{bux2014valid}. However, if such an algorithm is only tested in isolation as a unit test, the test lacks credibility since it does not reflect the real-world sequence of complex states. For instance, GC logic activates only if a certain precondition (e.g., a threshold of invalid pages) is met. Neglecting these conditions during unit testing may result in a failure to adequately verify its intended functionality \cite{karlsson2019exploratory}. Ultimately, these routines can only be thoroughly tested once sufficient I/O traffic has established the necessary internal states.

\paragraph{\textbf{Coverage-Based Fuzzing (CGF) and the Use of State Information}}
Coverage-based fuzzing (CGF) is widely employed in software testing due to its systematic exploration of program behavior. Tools such as AFL and LibFuzzer \cite{zalewski_afl,serebryany2015libfuzzer} generate and execute a vast number of inputs. Whenever a new code path is uncovered, they add the corresponding input to the corpus. This path-coverage feedback is primarily designed for programs with a deterministic relationship between inputs and outputs. Additional considerations are required for systems in which previous inputs affect current results, such as firmware. The need for state-aware fuzzing has grown in domains such as network protocols and embedded firmware. For example, AFLNet \cite{meng2025aflnet} targets network-server protocol implementations by tracing protocol states. Recent work leverages richer state feedback: PAVFuzz \cite{zuo2021pavfuzz} learns dependencies among automotive protocol fields to weight mutations, and StateFuzz \cite{zhao2022statefuzz} tracks Linux kernel globals to prioritize inputs that create new states. Both methods boost coverage by uncovering bugs that conventional fuzzers miss.

\paragraph{\textbf{Overcoming Nondeterministic I/O and Reinforcement Learning}}
Proposals have also been made to leverage NVMe I/O Determinism (IOD) to mitigate the unpredictability of internal maintenance tasks. The IODA technique \cite{li2023extending} introduces a predicted latency flag (PLIO) and a predicted time window (PLWin) method, reporting up to a 75-fold latency improvement by effectively controlling unpredictable delays caused by internal maintenance. However, it focuses on performance optimization, not on making the firmware state deterministic. Hence, its applicability to test reproducibility or maximizing coverage remains limited. Reinforcement learning has also been explored for SSD firmware testing. For example, GRAFT \cite{lee2023graft} models SSD firmware’s internal control flow as a graph, combining graph neural networks and offline RL to auto-generate test cases, achieving high coverage in short sequences. However, it assigns rewards based on discovering new basic blocks, making it difficult to intensively explore logic that manifests only after hundreds of thousands of I/Os. A lack of immediate rewards leads to long training gaps.

These approaches struggle to trigger SSD firmware routines that require extensive I/O accumulation. For example, StateFuzz observes state changes at a coarse granularity, whereas our approach directly leverages raw variable updates to drive long-term, threshold-based logic. Moreover, this technique applies not only to storage device firmware but to any embedded system governed by threshold-triggered operations.

\section{METHODS}
\subsection{Overview}
Figure 1 illustrates the fuzzing framework, which can be broken down into three key components:

\begin{itemize}[leftmargin=*, itemsep=1em, topsep=0em]
\item {\texttt{Coverage-based Fuzzing:}}
        Building on AFL++ \cite{fioraldi2020afl++} to track branching information in the SSD firmware under test, we treat any new code path's input (test sequence) as an interesting seed that is stored and mutated. We developed an additional function that forms command sequences by combining multiple NVMe CLI commands into a single “Test Sequence”.
\item {\texttt{State Data Monitoring:}}
        SSD firmware’s internal logic (e.g., GC) often activates only after crossing certain thresholds. Consequently, we monitor key variables in real time—such as victim line count, NAND block erase count, and NAND block invalid page count—via hooks in the FEMU emulator \cite{li2018case} or commands that provide direct access. If one or more state variables show a significant change from their previous values, it implies resource consumption or the triggering of certain logic. Accordingly, the corresponding Test Sequence is labeled a “Successful Test Sequence” and is specially tracked.
\item {\texttt{State Data-Aware Test Sequence Reuse:}}
        Sequences that trigger state changes are stored in a successful sequence pool. If these sequences continue to induce similar state changes upon replay, their weights are increased accordingly. Rather than blindly mutating new inputs, the fuzzer preferentially reuses sequences that previously caused changes in contexts similar to the current SSD state. Consequently, even in the absence of any immediate coverage increase, sequences that trigger critical state changes retain their value, enabling the fuzzer to quickly activate long-term routines (e.g., GC).
\end{itemize}

\setlength{\tabcolsep}{3pt}  
\renewcommand{\arraystretch}{1.0}  
\begin{table*}[htbp]
\centering
\caption{Performance summary of different fuzzers in SSD firmware fuzzing}
\label{tab:performance_combined}
\resizebox{\textwidth}{!}{%
\begin{tabular}{l*{5}{cc}ccc}
\toprule
 & \multicolumn{2}{c}{1st Trial} 
 & \multicolumn{2}{c}{2nd Trial} 
 & \multicolumn{2}{c}{3rd Trial} 
 & \multicolumn{2}{c}{4th Trial} 
 & \multicolumn{2}{c}{5th Trial} 
 & \multicolumn{3}{c}{Average} \\
\cmidrule(lr){2-3}\cmidrule(lr){4-5}\cmidrule(lr){6-7}\cmidrule(lr){8-9}\cmidrule(lr){10-11}\cmidrule(lr){12-14}
Fuzzer 
  & CMD\# & \begin{tabular}[c]{@{}c@{}}Crash/\\Hang\end{tabular} 
  & CMD\# & \begin{tabular}[c]{@{}c@{}}Crash/\\Hang\end{tabular} 
  & CMD\# & \begin{tabular}[c]{@{}c@{}}Crash/\\Hang\end{tabular} 
  & CMD\# & \begin{tabular}[c]{@{}c@{}}Crash/\\Hang\end{tabular} 
  & CMD\# & \begin{tabular}[c]{@{}c@{}}Crash/\\Hang\end{tabular} 
  & CMD\# & \begin{tabular}[c]{@{}c@{}}Time\\(min)\end{tabular} & \begin{tabular}[c]{@{}c@{}}Crash/\\Hang\end{tabular} \\
\midrule
\multicolumn{14}{l}{\textbf{Write/Read Only}} \\
Random Fuzzer 
  & 533,993,541 & 2/0 
  & --- & --- 
  & --- & --- 
  & --- & --- 
  & --- & --- 
  & 533,993,541 & 4320 & 2/0 \\
AFL++         
  & 7,408,791  & 9/4 
  & 19,706,163 & 10/5 
  & 27,588,652 & 10/7 
  & 9,402,423  & 10/8 
  & 53,096,280 & 11/5 
  & 23,440,462 & 419.2 & 10/5.8 \\
Proposed Method 
  & 1,378,339  & 11/5 
  & 8,781,059  & 8/7 
  & 2,673,291  & 8/5 
  & 8,400,502  & 9/8 
  & 17,085,887 & 9/4 
  & 7,663,816  & 143   & 9/5.8 \\
\midrule
\multicolumn{14}{c}{\textbf{Proposed Method's Command Execution 67.3\% Lower, Time 65.9\% Lower than AFL++}} \\
\midrule
\multicolumn{14}{l}{\textbf{6 Types of I/Os}} \\
AFL++         
  & 55,688,797 & 18/7 
  & 170,648,503 & 18/7 
  & 41,610,075 & 20/6 
  & 45,524,401 & 17/11 
  & 85,838,539 & 18/10 
  & 79,862,063 & 695.4 & 18.2/8.2 \\
Proposed Method 
  & 5,459,667  & 14/3 
  & 15,322,687 & 20/10 
  & 21,741,295 & 17/10 
  & 9,970,934  & 20/8 
  & 25,041,211 & 18/8 
  & 15,507,159 & 175.8 & 17.8/7.8 \\
\midrule
\multicolumn{14}{c}{\textbf{Proposed Method's Command Execution 80.6\% Lower, Time 74.7\% Lower than AFL++}} \\
\bottomrule
\end{tabular}%
}
\end{table*}

\subsection{Implementation}
Below is a summary of the actual implementation steps, explaining how the proposed technique generates, evaluates, and reuses Test Sequences:
\begin{enumerate}[leftmargin=*, itemsep=0em, topsep=0em]
  \item    Initial Test Input and State Logging:
        We use a simple NVMe CLI command (e.g., a write command) as the initial seed. SSD state variables are logged in shared memory, and initial values are recorded.
  \item     Coverage-Guided Test Input Generation:
        We modified AFL++ so that each mutated byte array is converted into an NVMe command sequence limited to 100 commands, controlling memory usage and overhead. To evaluate the impact of this limit, we also ran experiments with 500-command sequences; the results are presented in Section 6.
  \item     Check State Data Changes:
        If any significant change is detected after execution, indicating resource consumption or the triggering of relevant logic, we store the recent Test Input set as a “Successful Test Sequence”.
  \item     Update Weight:
        Each state variable is weighted inversely proportional to its observed change frequency—variables that change infrequently receive higher weights. Whenever a test sequence causes a variable to cross its predefined threshold, we increase that sequence’s weight by an amount proportional to the variable’s weight. Consequently, sequences that successfully drive threshold-crossing changes in more stable state variables are given higher selection priority.
  \item     Sequence Reuse Analysis:
        On each iteration, the fuzzer compares the SSD’s current I/O state profile against those of previously successful, state-changing sequences. It then computes a similarity score and replays the sequence most likely to reproduce the same state transition.
  \item     Execute Test Input or Reuse:
        The chosen test sequence is executed. Afterward, we return to Step (3) to check for state data changes and repeat the loop.        
\end{enumerate}

\subsection{Rationale: Case-Based Reasoning and State Data Usage}
\paragraph{Why Use State Variables?}
    Many SSD firmware routines require prolonged I/O accumulation before activation, making random exploration inefficient. Detecting changes in internal state variables provides valuable signals that a threshold is approaching, enabling more effective test sequence selection \cite{myung2022mundofuzz}. Such state information is often accessible via vendor-specific commands, supporting both grey-box and real-world testing scenarios.

\paragraph{Necessity of Test Sequence Reuse}
    SSD address mapping and asynchronous scheduling can cause significant nondeterminism, so we adopt a Case-Based Reasoning (CBR) \cite{kolodner2014case} approach to support efficient sequence reuse.:
\begin{itemize}
\item {\texttt{Retrieve}}:
        From our pool of previously successful sequences, choose those whose stored I/O state profiles most closely align with the current SSD conditions.
\item {\texttt{Reuse}}:     
        Replay or lightly mutate the retrieved sequence, prioritizing command patterns that historically drove meaningful state changes.
\item {\texttt{Revise}}:     
        If replay does not yield the expected change (e.g., due to asynchronous noise), we demote that sequence’s weight or reorder its commands to suppress irrelevant operations.
\item {\texttt{Retain}}:     
        Only sequences that consistently induce the target threshold-crossing are kept; non-reproducible sequences are discarded to prevent noise accumulation.
\end{itemize}
This CBR framework leverages reliable past instances to mitigate asynchronous events and invalid inputs, actively accelerating long-term, threshold-driven firmware routines.       

\subsection{Test Sequence Ontology}
Simply listing commands makes it hard to see which part of a sequence drives state changes. Each test sequence comprises a combination of commands that records which commands are executed, where they are applied, and how many times. To formalize this structure and enable systematic matching and reuse, we adopt a Testcase Ontology \cite{popereshnyak2019modeling,li2015use}. For the Precondition, the categories Hot, Warm, and Cold represent logical address regions that are frequently accessed. This categorization was introduced to capture the accumulated I/O state of the SSD, thereby providing a useful metric for gauging the firmware’s internal I/O behavior \cite{song2023f2fs}.
This is a standardized structure for a single Test Sequence:
\begin{enumerate}[label=\textbf{\arabic*.}, leftmargin=2.1cm, labelwidth=2cm, align=left]
  \item[ID:] Unique identifier
  \item[Name:] Test name
  \item[Precondition:] Current SSD’s accumulated I/Os
  \item[Input:] Command parameters (e.g., read size, range)
  \item[Operation:] Frequency of each command type
  \item[Expectation:] Success/failure for each command, return codes
\end{enumerate}

\begin{listing}[htbp]
\begin{MyVerbatim}
ID: (1)
Name: (Successful Test Sequence 1)
Precondition: (Cold, Cold, Warm, Warm, Hot, ...)
Input: (Hot=1, Warm=4, Cold=5)
Operation: (Write=4, Compare=1, Flush=4, Read=1)
Expectation: (Success=6, Fail=4)
\end{MyVerbatim}
\caption{Test Sequence with 10 NVMe CLI commands}
\label{lst:testsequence}
\end{listing}

To capture finer-grained precondition states, we partition each logical region’s I/O-accumulation range into ten equal bins and map those bins to Cold, Warm, or Hot categories (e.g., bins 1–5 → Cold, 6–9 → Warm, 10 → Hot). We then encode the SSD’s current profile as a three-valued vector over these ten segments. To select a sequence for replay, we compute the Manhattan distance between this runtime vector and each stored precondition vector, directly leveraging the raw category assignments rather than absolute magnitudes. By replaying the sequence with the smallest distance, we execute the test sequence under conditions similar to those that originally induced state changes, thereby rapidly activating threshold-based routines.

\section{EXPERIMENTAL SETUP}
\subsection{Research Questions (RQ)}

This work answers the following two RQs to evaluate the performance and utility of the state data-aware fuzzer:
\subsubsection{RQ1. How efficiently does it achieve coverage and detect defects?}

    We measure the number of commands and time needed to reach full coverage of I/O-related code. For Write/Read-only tests, we instrument the FTL (address mapping, GC) and related error-handling routines triggered by these operations. We define “100\% I/O coverage” as executing all of these instrumented basic blocks. Code paths requiring other commands lie outside this scope, as they are unreachable in this setup. Also, we compare the number of crashes or hangs occurrences recorded by AFL++ to determine whether the state-aware fuzzer uncovers additional defects. (Duplicate crashes/hangs are counted as one.)

\subsubsection{RQ2. Does it remain effective when additional commands beyond basic I/O are tested?}

    We expand the input set to six I/O commands (Write, Read, Compare, Flush, Write-Zeroes, Write-uncorrectable) to evaluate coverage efficiency.
    We also analyze which commands appear most frequently in test sequences, and whether there is a difference between the commands used in early and later stages of fuzzing.

\subsection{Experiment Environment and Benchmark}

We evaluated the proposed method on FEMU \cite{li2018case}, an NVMe SSD emulator that includes typical SSD internals such as an FTL and a GC algorithm. To collect precise coverage data, we instrumented FEMU by compiling it with AFL++’s LLVM LTO toolchain (afl-clang-lto/afl-clang-lto++) and enabled Clang’s built-in profile-guided coverage instrumentation. We then ran AFL++ together with an NVMe Command Line Interface (NVMe CLI) harness to send I/O commands for fuzzing.

We compared the following three fuzzing strategies:
\begin{description}[leftmargin=0em, itemsep=0em, topsep=0em]
  \item[Random Fuzzer:] We implemented a baseline fuzzer that issues NVMe I/O commands via the NVMe CLI, with both the Logical Block Address and the Number of Blocks parameters drawn uniformly at random for each request. After three days of testing, no additional coverage was observed, and even restricting operations to Write and Read did not enable further exploration. Therefore, only a single trial result was recorded.
  \item[AFL++:] A state-of-the-art coverage-based fuzzer (AFL++) in default configuration, unaware of any state information—if coverage increases, the input sequence is added to the corpus.
  \item[Proposed Fuzzer:] Our approach leverages both coverage and state changes as feedback, reusing successful sequences.
\end{description}

All fuzzers started from the same initial seed (a simple Write command), and FEMU was reinitialized before each run. We also prevented reuse of previously discovered coverage across campaigns.
\subsection{Evaluation Metrics}
\begin{description}[leftmargin=0em, itemsep=0em, topsep=0em]
  \item[Command Efficiency:] The number of commands (and total time) needed to attain 100\% I/O-relevant coverage.
  \item[Crash/Hang Count:] The number of fatal exceptions (crashes) or unresponsive states (hangs) observed during fuzzing. Repeated triggers of the same Crash/Hang root cause are counted as one. We use AFL++’s reporting feature to log these events.
\end{description}

\section{RESULTS}
Table 1 summarizes the main performance metrics for each fuzzer, and we analyze the key findings for each RQ below.
\subsection{RQ1: Coverage Efficiency and Defect Detection}

    \paragraph{Coverage vs. Command Count.} The proposed state data-aware fuzzer reached 100\% I/O coverage of the SSD firmware code with about 7.7 million commands, whereas AFL++ required about 23.4 million. In other words, our method used 67.3\% fewer commands while achieving the same coverage. Meanwhile, the random fuzzer ran more than 500 million commands over 3 days and still failed to hit 100\% coverage, plateauing at around 89\%.
    In FEMU, the GC routine is activated once victim line count — the number of flash-memory lines marked as candidates for reclamation — increases to 190. AFL++ did not accelerate this reduction effectively, whereas the state-aware fuzzer quickly forced victim line count increase and triggered GC. The random fuzzer never reached that threshold or changed this value during our observation period. Figure 2 shows how these metrics changed over time for each fuzzer.

\begin{figure}[h]
  \centering
  \includegraphics[width=\linewidth]{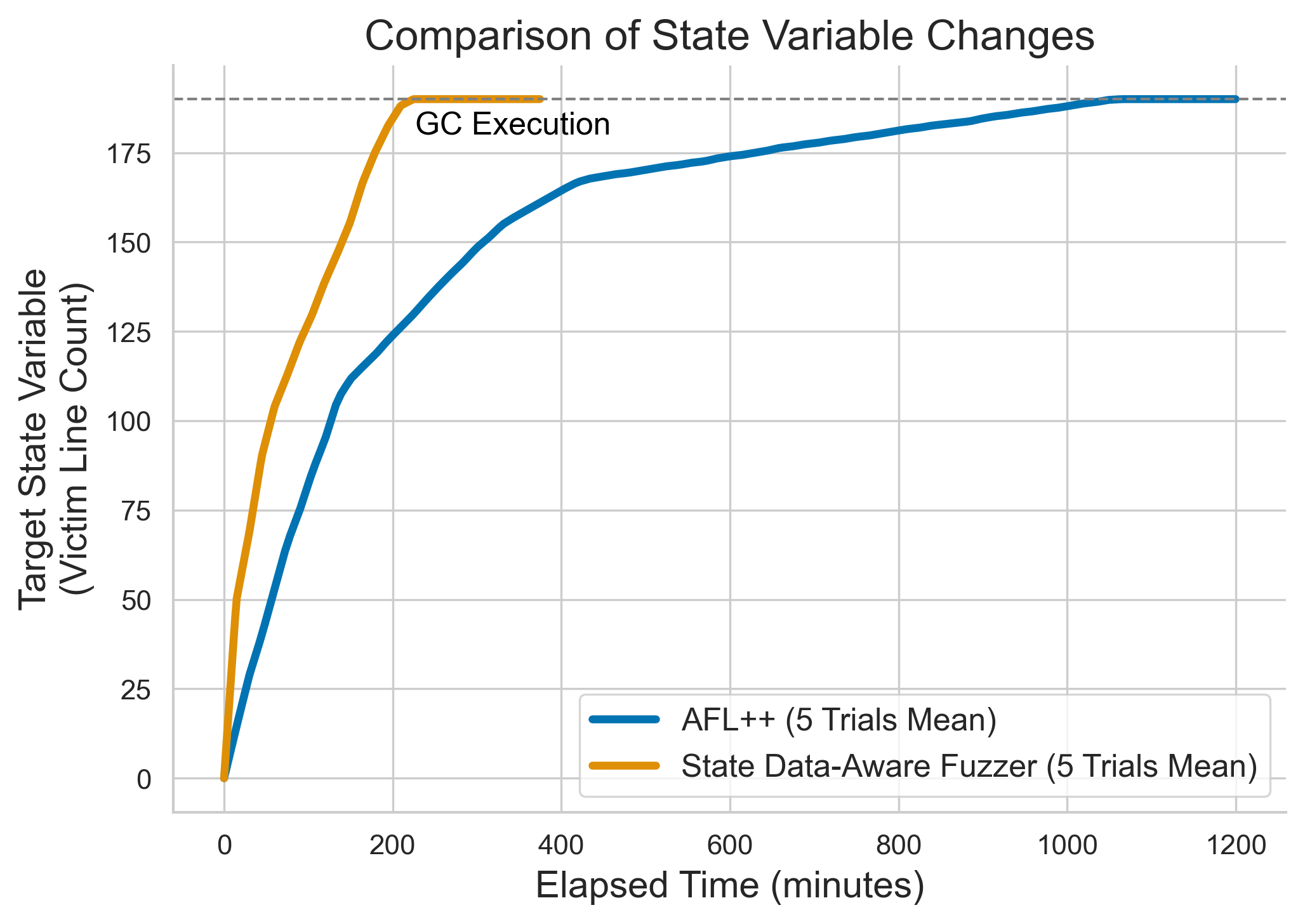}
  \caption{State value changes. When the value reaches 190, internal firmware maintenance logic is triggered.}
  \Description{State Value Changes}
\end{figure}

    \paragraph{Bug Detection.} On identical hardware, our state-aware fuzzer required about 65\% less time, and from a defect detection perspective, it discovered 94\% of the crashes and hangs found by AFL++ (96\% in the extended I/O case). It detected slightly fewer bugs because we ended each run upon reaching 100\% I/O coverage, using only about one-third of AFL++’s total fuzzing time. Assuming the same number of commands, tracking states and managing sequences incurs a 16\% overhead relative to AFL++; however, since state-aware fuzzing needs fewer commands overall to reach full coverage, the total fuzzing time was roughly one-third that of AFL++, indicating little bottleneck in the fuzzing loop.

\subsection{RQ2: Expanded I/O Command Experiments}

\begin{figure}[h]
  \centering
  \includegraphics[width=\linewidth]{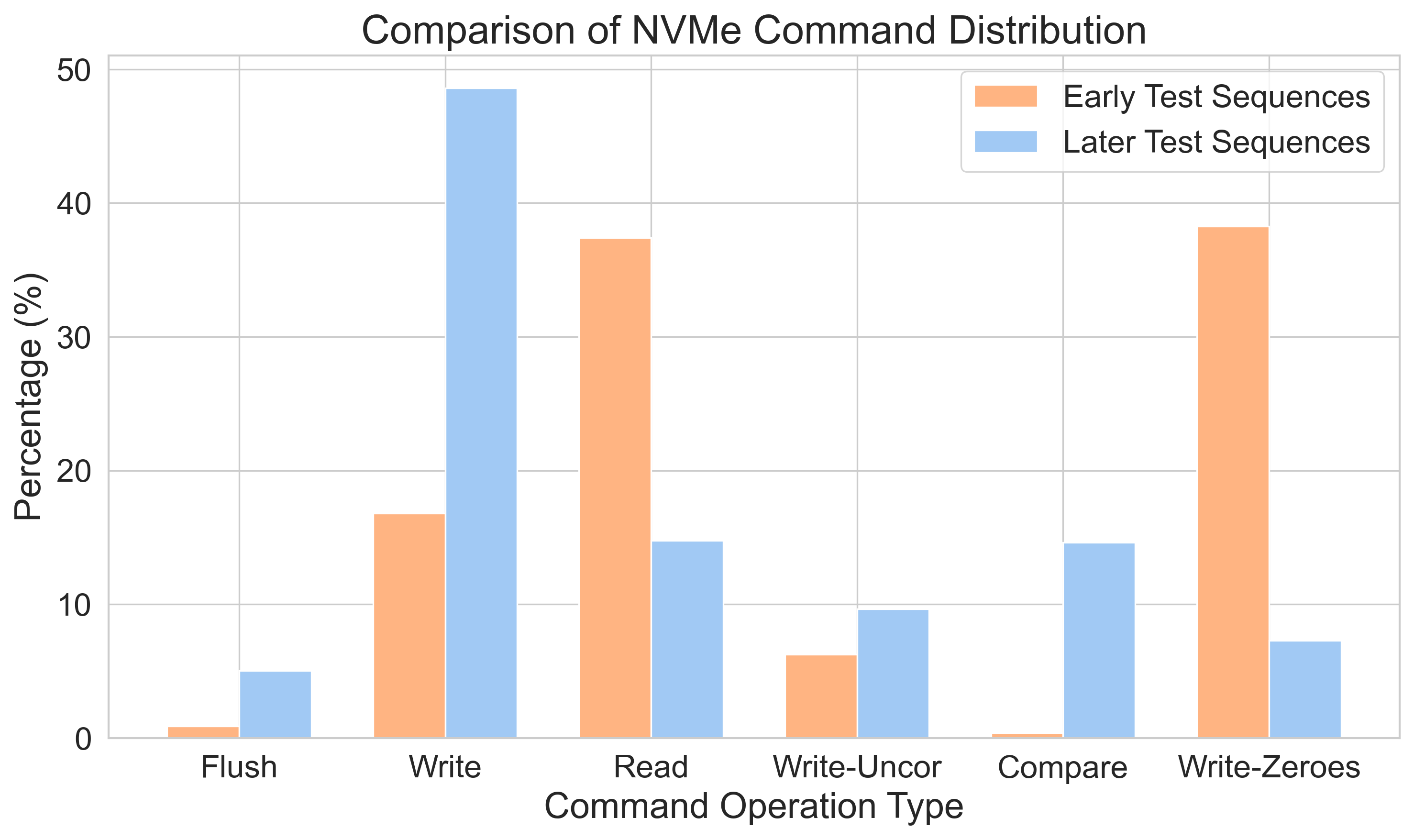}
  \caption{Distribution of Command Operations. Later Test Sequences have more Write and Flush Commands.}
  \Description{Distribution of Command Operations}
\end{figure}

Upon expanding the input space, our results were particularly compelling. In addition to the standard Write and Read commands, we incorporated Flush, Compare, Write-Uncorrectable, and Write-Zeroes operations. All fuzzers exhibited a slower exploration rate because some commands do not directly impact the flash layer, thereby yielding less effective inputs and enlarging the search space. Nevertheless, the state data-aware fuzzer maintained a high level of efficiency. For example, since both Write and Flush directly affect the FTL, our fuzzer quickly recognized this impact and adjusted the command distribution by increasing the usage of these operations—a trend that Figure 3 clearly demonstrates. As a result, we achieved overall I/O coverage using approximately 80.6\% fewer commands than AFL++. In terms of bug detection, the expanded set of I/O commands uncovered 10 additional crashes, with the state data-aware fuzzer detecting 96\% of these within only one quarter of the time required by the AFL++ experiment. These findings suggest that state feedback remains effective across broader input spaces by learning and exploiting the impact of each command on internal states. Moreover, because these commands are genuine NVMe SSD operations, they confirm that the proposed method is valid in realistic I/O environments similar to those used by real users.

\section{DISCUSSION}
\sloppy
Our experiments confirm that incorporating state awareness into SSD firmware fuzzing significantly improves the fuzzer’s ability to detect and drive threshold-based routines. By monitoring internal variables, the fuzzer recognizes when the firmware is nearing threshold-triggering states and adapts sequence selection accordingly—capabilities that traditional coverage-only fuzzers lack. Although nondeterministic I/O can introduce noise, our case-based reuse of sequences under matching precondition profiles provides a robust guiding signal. This approach generalizes to any system with observable internal states.

Furthermore, we evaluated the impact of increasing the test sequence limit from 100 to 500 commands. Over five independent runs under the 500-command limit, achieving the same I/O coverage required on average 7,527,963 commands and 134 minutes. These results correspond to 98.2\% of the command count and 95.4\% of the runtime relative to the 100-command experiments, indicating that longer sequences can partially reduce overall effort. While finding the most efficient sequence length is a promising idea, its effectiveness is likely to vary with different firmware implementations and maintenance algorithms; tuning this parameter represents a valuable avenue for future work.

Future work will explore statistical filtering to further reduce noise, snapshot-based execution to handle nondeterminism more systematically, and adaptive optimization of test-sequence length tailored to specific firmware and algorithmic characteristics.

\section{THREATS TO VALIDITY}
\paragraph{Internal Validity} This study monitored only a subset of firmware variables (e.g., victim line count, invalid page count) closely tied to threshold-based logic. If other critical variables were omitted, certain code paths or maintenance routines might remain insufficiently tested. We plan to expand our monitoring scope or employ automated analyses to identify additional variables.

Our approach reuses sequences that induce state changes to focus on routines requiring prolonged I/O accumulation. Although this could potentially overlook simpler paths, coverage-based fuzzers typically explore those paths early on, making the overall impact less significant. Nevertheless, additional validation across diverse scenarios is required to ensure that no critical logic is omitted.

Lastly, fuzzing relies on mutational processes, so outcomes can vary with random seeds and system load. Although we repeated each experiment five times and reported the average, we cannot completely rule out variations due to limited repetition or differing test environments.
\paragraph{External Validity} We evaluated our approach using the QEMU-based FEMU emulator for SSD firmware, which may diverge from real hardware in terms of timing, parallelism, and scheduling. Hence, additional validation on physical SSDs is necessary to confirm consistent performance gains.

\section{CONCLUSION AND FUTURE WORK}
This paper demonstrates that adding state data-aware fuzzing to SSD firmware testing enables rapid coverage under nondeterministic I/O and threshold-triggered maintenance. By reusing successful test sequences to activate these routines, our method matches the coverage of state-of-the-art fuzzers (e.g., AFL++) with approximately 67.3\% fewer commands, while still reliably detecting crashes and hangs.

In the future, we plan to explore automated selection of internal state variables, practical testing on real SSD hardware, and improvements via distributed fuzzing to enhance exploration even further. Ultimately, these directions aim to make fuzzers more aware of and able to leverage internal states, thereby improving reliability across a broad range of complex embedded and storage systems.

\section{DATA AVAILABILITY}
The raw data are available in Zenodo at \url{https://doi.org/10.5281/zenodo.15285200} (CC-BY 4.0)

\bibliographystyle{ACM-Reference-Format}
\bibliography{MyReference}


\end{document}